\documentclass[onecolumn,prb,aps,superscriptaddress,showpacs]{revtex4}
\usepackage{amsmath}
\usepackage{graphicx}
\usepackage[version=3]{mhchem}

\begin{document}
\title{Topological bands in one-dimensional periodic potentials}
\author{Yi Zheng}
\affiliation{Department of Physics, Beijing Normal University, Beijing 100875, China}
\author{Shi-Jie Yang\footnote{Corresponding author: yangshijie@tsinghua.org.cn}}
\affiliation{Department of Physics, Beijing Normal University, Beijing 100875, China}
\affiliation{State Key Laboratory of Theoretical Physics, Institute of Theoretical Physics, Chinese Academy of Sciences, Beijing 100190, China}
\begin{abstract}
We study the properties of the quantum states in the one-dimensional system with a shifted periodic potential in both the discrete model and the continuous model. With open boundary conditions, the edge states appear in the energy gaps which indicate non-trivial topological structures. The Chern numbers with respect to the Bloch vector and the potential shift angle are computed. In the limit of the continuous model, the Chern number of each band is exactly one. We demonstrate the particle number pumped by the adiabatically shift of the potential is directly related to the topological invariants.

\end{abstract}
\maketitle
\section*{1. Introduction}
Condensed matter physics has been concentrating largely on finding different states of matters. One of the greatest recent achievements is the discovery of new form of states by applying the concept of topology which distinguish the states from the others characterized by Landau's symmetry breaking theory\cite{1wen}. This progress starts from the finding of the quantum Hall effect (QHE) in the two-dimensional (2D) electron gas\cite{2Klitzing}, and then immediately causes wide interests\cite{3qi and zhang}. It was found that the Kubo's formula for quantized Hall conductance lead to the topological invariant\cite{4Thouless,5Kane}. Different states with distinguished conductance have the same symmetry and hence cannot be classified by the broken symmetries.

While the large magnetic field in the QHE system limits its possible applications, scientists proposed a new kind of insulating state without the magnetic field\cite{6Kane,7Ber}. The spin up and the spin down electrons have the same Hall conductance but with opposite signs leading to the so-called quantum spin Hall effect as well as the 2D topological insulator, owing to the spin-orbit couplings. Experimentally, this new state of matters have been observed and well studied in $TeHg$ quantum well as well as in alloys like $Bi_{1-x}Sb_x$ and $Bi_2Te_3$\cite{8,9,10}. On the other hand, the development of optical lattices in cold atom physics has also stimulated the exploration for topological states\cite{11,12}.

The 2D topological insulators have been extended to three-dimensional systems. On the other hand, the 1D systems are usually supposed to be topologically trivial. Because topology arises as the Brillouin zone (BZ) forms a torus and the wave vectors in both the $x-$ and $y$-direction are needed. In recent years, some researchers have shown nontrivial topological properties in 1D systems by introducing a system parameter as a virtual dimension\cite{13}. It has been proved that these properties are relevant to higher dimensional space\cite{14}. For example, a shift $\delta$ of the periodic potential in the 1D Harper model is introduced to make the Hamiltonian periodic in the parameter space. This potential shift plays the role of the Bloch vector in the $y$-direction.

As the Hamiltonian is periodic for both ${k}_{x}$ and $\delta$, the effective 2D BZ forms a torus. In the QHE, the existence of edge states manifests the topological order. For 1D case, the edge states also appear in the band gap in which the electrons are localized at the edge. In the present work, we study the topological properties in 1D periodic potentials, in either discrete or continuous systems. We show that the nonzero Chern numbers for the energy bands. We also demonstrate that the particle pumping driven by adiabatic shift of the periodic potential is quantized and is intimately related to the topological Chern number.

The paper is organized as follows: In Sec.II we present the model and display the band structures, both in the discrete model and in the continuous model. In the open boundary condition there are edge states present in the energy gaps. In Sec.III we compute the Chern numbers of the individual energy band. In Sec.IV, we demonstrate the quantized electron pumping, which is related to the topological invariants. A brief summary is included in Sec.V.

\section*{2. Band structure}
We start from the 1D system with a trigonometric potential $V(x)=V_0\cos(2\pi\alpha x+\delta)$, where $\alpha=p/q$ ($p$ and $q$ are co-prime integers) is a rational number which defines the period, it roots originally from 2-D system with vertical magnetic field, and can be understood as the ratio of flux through a lattice cell to one flux quantum\cite{15}. $\delta$ represents a shift of the periodic potential with $V(\delta+2\pi)=V(\delta)$. We take $0\leq\delta<2\pi$. The stationary Schrodinger equation reads $[-\frac{1}{2}\partial_x^2+V(x)]\phi= {E}\phi$. By considering the Bloch state $\psi=e^{ik x}\phi$ for $\left|k\right|\le\pi/q$, the finite differential equation is
\begin{equation}\label{a}
-\frac{e^{ik\cdot\Delta x}\psi_{j+1}+ e^{-ik \cdot\Delta x}\psi_{j-1}-2\psi_j}{2(\Delta x)^2}+V_j\psi_j = E_n\psi_j.
\end{equation}
It can be rewritten as
\begin{equation}\label{b}
-(e^{ik\cdot\Delta x}\psi_{j+1}+e^{-ik\cdot\Delta x}\psi_{j-1})+V_j\psi_j=\varepsilon_n\psi_j,
\end{equation}
where ${\varepsilon_n}=2{(\Delta x)^2}{E_n}-2$. Equations (\ref{b}) share the same ingredient with the discrete Harper equations in the tight binding model as $\Delta x=1$: $-(u_{j+1} + u_{j-1}) + V_j u_j=\varepsilon_n u_j$ by substitution of $u_j=e^{ikx_j}\psi_j$ at the $j$-th site. It can be expressed as the matrix equation $M\psi=\varepsilon_n\psi$, where $M$ is a $q\times q$ matrix which is to be diagonalized in the eigenvalue problem. The energy spectrum splits into $q$ bands.

\begin{figure}[h]
\begin{center}
\includegraphics*[width=8.5cm]{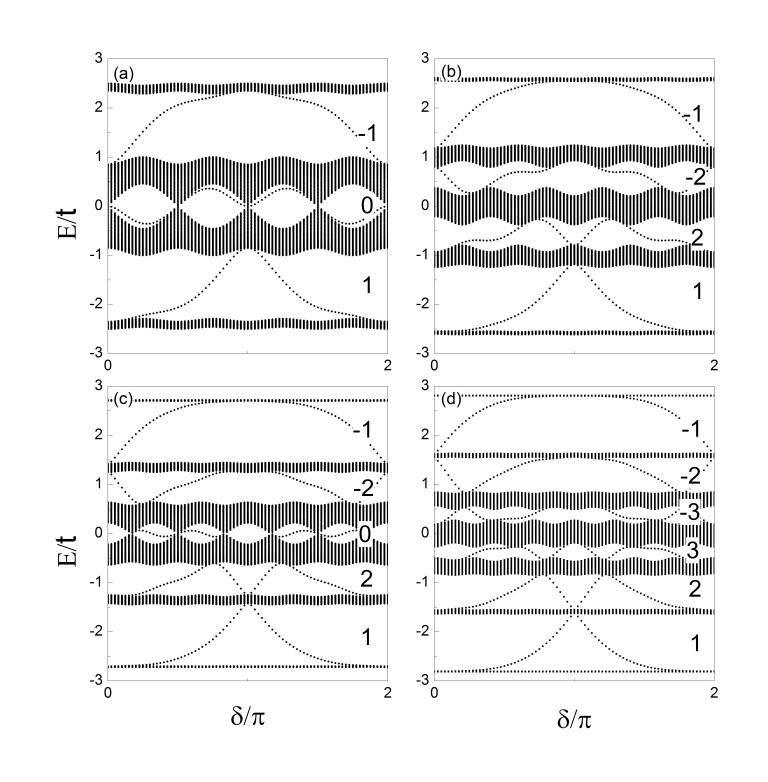}
\end{center}
\caption{The energy spectrum with open boundary condition for $p=1$: (a) $q=4$, $N=40$. (b) $q=5$, $N=40$. (c) $q=6$, $N=30$. (d) $q=7$, $N=30$. In all cases $V_0=1.5$. The states in the gap are edge states. The numbers in each gap are the overall Chern number of the bands below the Fermi level (see Table I).}
\end{figure}

As to the periodic boundary condition, the energy spectrum can be depicted in the $E-k$ diagram at a fixed $\delta$. For the open boundary condition, the Bloch vector $k$ is no longer a good quantum number. Instead we draw the $E-\delta$ spectra. The shift parameter $\delta $ plays the role similar to that of the Bloch vector $k$. Figure 1 and Fig.2 display the spectra of the discrete model ($\Delta x=1$) of various finite $q$ with $p=1$ and $p=2$, respectively. The number of periods $N$ determines the number of states in a single band. We note that edge states appear in the band gap, which indicate the electrons are localized at the edges\cite{13}. The numbers denoted in the gaps are the overall Chern numbers of the lower bands as the Fermi level falls in the gaps. The details are discussed in the next section.
\begin{figure}[h]
\begin{center}
\includegraphics*[width=8.5cm]{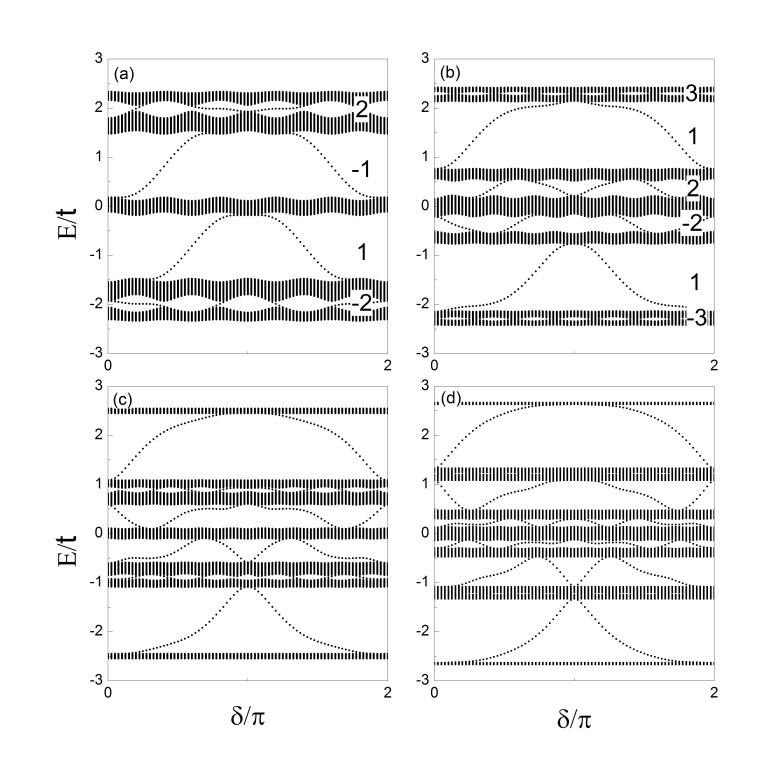}
\end{center}
\caption{The same as in Fig.1 for $p=2$: (a) $q=5$, $N=40$. (b) $q=7$, $N=40$. (c) $q=9$, $N=25$. (d) $q=11$, $N=20$.}
\end{figure}

For a continuous system of spatial period $d$, we come back to the equation (\ref{a}) with the  correspondence of $q=d/\Delta x\to\infty$ as $\Delta x \to 0$. There are infinite number of bands. Only lower bands are of interested for a real system. Figure 3 show two examples of continuous periodic potentials of (a) $V(x)=V_1\cos(2\pi x/d+\delta)$ and (b) $V(x)=V_1\cos(2\pi x/d+\delta)+V_2\cos(4\pi x/d+2\emph{}\delta)$, respectively. One (Fig.3(a)) or two (Fig.3(b)) gaps are opened in the two cases. The edge states in the gap indicate nontrivial topology in the system. The numbers indicated in the gaps are the overall Chern numbers. An arbitrary periodic potential $V(x+d)=V(x)$ can be expressed in the Fourier series: $V(x)=\sum_{j=1}^\infty [V_{1j}\cos(2\pi jx/d)+V_{2j}\sin(2\pi jx/d)]$. In the first order approximation, the width of the gap between the $j$-th and the $(j+1)$-th bands is proportional to the values of the coefficients $V_{1j}$ and/or $V_{2j}$. The Chern number of each band is exactly equal to one. We conclude that the band structure in a continuous potential generally exhibits topological characteristics. If ${V_j}$ are absence, the gap is too small and the Chern number becomes meaningless.

\begin{figure}[t]
\begin{center}
\includegraphics*[width=8.5cm]{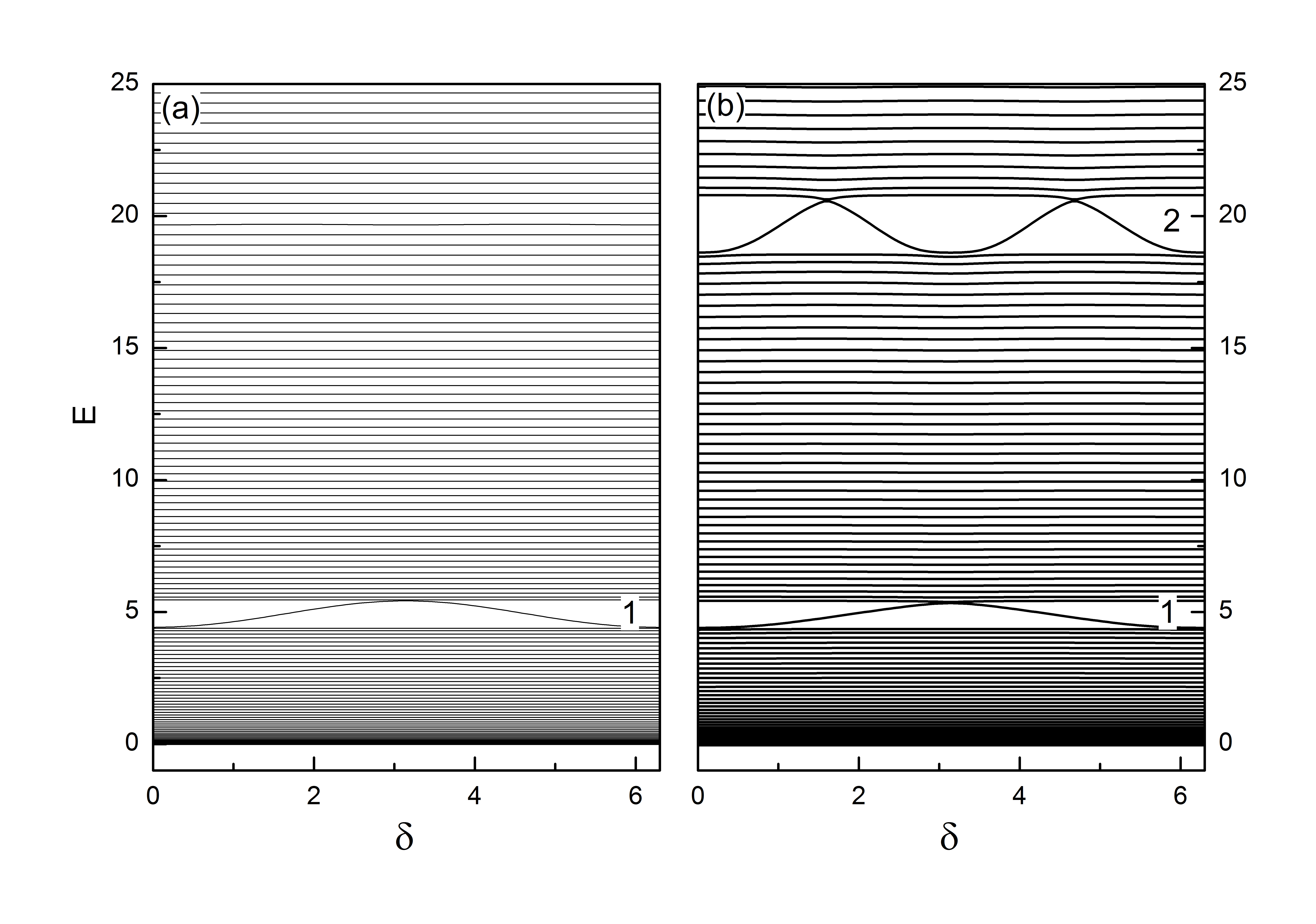}
\end{center}
\caption{The energy spectra in the continuous potential model under the open boundary condition: (a) $V=V_1\cos(2\pi x/d+\delta)$. (b) $V=V_1\cos(2\pi x/d+\delta )+V_2\cos(4\pi x/d+2\delta)$. $N=50$ and $V_1=V_2=1.5$. As $N\to \infty$, the bands become continuous. The edge states appear in the gaps. The numbers in each gap are the overall Chern number of the bands below the Fermi level.}
\end{figure}

\section*{3. Topological invariant}
So far we have an effective 2D BZ with respect to the Bloch vector $k$ and the potential shift $\delta$. It forms a $T^2$ torus since the system is periodic in the two directions. The topological structure is characterized by the topological invariant or the Chern number of the occupied bands, which is the total Berry flux ${F_n}$ in the BZ. For convenience, we denote $k$ and $\delta$ by ${k_1}$ and ${k_2}$, respectively. The Chern invariant can be computed via
\begin{equation}
c_n=\frac{1}{2\pi}\int_{T^2}{d^2k}F_n,\label{x}
\end{equation}
where $F_n=\nabla\times{A_n}$ is the field strength and $A_n= i\left\langle{u_n}\right|\left.{\nabla_k}\right|\left. {{u_n}} \right\rangle $ is the Berry connection which contributes to the Berry phase\cite{16}.

To compute the Chern invariant for each isolated band, we employ the method proposed in Ref.[\onlinecite{17}]. The numerical results are shown in Table \uppercase\expandafter{\romannumeral1} for $\alpha =1/q$ and $\alpha = 2/q$. The Chern numbers differs only in sign for $\alpha =p/q$ and $\alpha=1-p/q$ cases. So for a given $q$, say $q = 7$, only fewer of $\alpha$ like $1/7$, $2/7$ and $3/7$ are relevant.
\begin{table}[h]
\begin{tabular}{|c|c|c|c|c|c|c|c|c|}
\hline
$p/q$ & 1/3 & 1/4 & 1/5 & 1/6 & 1/7 & 2/5 & 2/7 & 2/9 \\
\hline
      &    &    &    &    &    &    &    & -4 \\
      &    &    &    & 1  & 1  &    & -3 & 5  \\
      &    & 1  & 1  & 1  & 1  & -2 & 4  & -4 \\
      & 1  & -1 & 1  & -2 & 1  & 3  & -3 & 5  \\
$c_n$ & -2 & -1 & -4 & -2 & -6 & -2 & 4  & -4 \\
      & 1  & 1  & 1  & 1  & 1  & 3  & -3 & 5  \\
      &    &    & 1  & 1  & 1  & -2 & 4  & -4 \\
      &    &    &    &    & 1  &    & -3 & 5  \\
      &    &    &    &    &    &    &    & -4 \\
\hline
\end{tabular}
\caption{The Chern number of each energy band for different parameter $\alpha=p/q$. Each column has $q$ bands. The numbers in the gaps of Fig.1 and Fig.2 are the sum of Chern numbers of the bands below the Fermi level. For the continuous potential systems which correspond to $q\to \infty$, the Chern number of each band is uniquely one.}
\end{table}

When the Fermi level falls in the $i$-th gap, there are $i$ occupied bands. The overall topological invariant for the filled bands is summation of the Chern number of each band as shown in Fig.1 and Fig.2. Chern number of the total bands vanishes due to mathematical concern\cite{18}. For the bands of even $q$, we get a zero Chern number as half of total bands are filled, indicating that the winding numbers of the phase of the transition function in the vector space spanned by orthogonal eigenstates are not well defined. It coincides with the spectra in which the considered gap with null Chern number closes up at some points. If we skip the gap with degenerate points and include next-nearest-neighbor hopping, we will get a well-defined winding number\cite{19}.

For continuous periodic model, the particle-hole symmetry is broken and the Chern number of each band equal to one since $q\to \infty$. The overall topological invariant corresponds to an additive integer, analogous to the Landau levels in the QHE systems.

\section*{4. Particle pumping}
The topological structure of this 1D system is caused by the introduced potential shift $\delta$, which makes the Hamiltonian periodic in the parameter space. Suppose that $\delta$ adiabatically changes with time as $\delta(t + T)=\delta(t)+2\pi$, then the Hamiltonian $H$ varies periodically with time as $H(t + T)= H(t)$. It is natural to relate the $\delta$ shift with the changes produced by adiabatic variations of the magnetic flux which has been well studied before\cite{20}. The time-dependent Schrodinger equation can be solved in the spirit of the adiabatic condition. In the first order perturbation, the instantaneous normalized eigenstate can be obtained and the average velocity of the particles is\cite{21}
\begin{eqnarray}
v_n(k) &=&\frac{\partial \varepsilon_n(k)}{\hbar \partial k} - i\sum_{n' \ne n}\left\{ {\frac{{\left\langle {{u_n}} \right|\partial_k H\left|u_{n'} \right\rangle \left\langle u_{n'}
\left |\right.
\partial_t{u_n} \right\rangle }}{\varepsilon_n - \varepsilon_{n'}} - c.c.} \right\} \nonumber\\
&=& \frac{\partial \varepsilon_n(k)}{\hbar \partial k} - i\left[ \left\langle\partial_k{u_n}
\left | \right.
\partial_t{u_n}\right\rangle -\left\langle \partial_t{u_n}
\left | \right.
\partial_k{u_n}\right\rangle \right].\label{c}
\end{eqnarray}

The derivations have made use of the facts that $\left\langle {u_n} \right|\partial_k H\left| u_{n'} \right\rangle = (\varepsilon_n - \varepsilon_{n'})\left\langle\partial_k{u_n}
|{u_{n'}} \right\rangle $ and $\sum_{n'}\left| u_{n'}\right\rangle \left\langle u_{n'}\right| = 1$. The adiabatic current is the integration of ${v_n}(k)$ over the BZ, in which the zeroth order term vanishes. The total particles pumped in the process is given by
\begin{eqnarray}
Q_n&=&i\int_0^T {dt} \int_\textrm{BZ}\frac{{dk}}{2\pi} \left[ \left\langle {{\partial _t}{u_n}|{\partial_k}{u_n}} \right\rangle - \left\langle \partial_k{u_n}|\partial_t{u_n} \right\rangle  \right],\nonumber\\
&=&i\int_0^{2\pi } {d\delta } \int_\textrm{BZ}{\frac{{dk}}{{2\pi }}} \left[ {\left\langle \partial_\delta{u_n}|\partial_k{u_n} \right\rangle - \left\langle \partial_k{u_n}|\partial_\delta{u_n} \right\rangle } \right].
\label{e}
\end{eqnarray}
The second equivalence establishes by taking account of the relation between $\delta $ and $t$. This formulism is exactly the Chern number expression in Eq.(\ref{x}) and is irrelevant of the amplitude of driving force. It demonstrated that the adiabatic pumping is quantized and the number of pumped particles is equal to the topological invariant.

The adiabatic pumping of particles stems from the charge pumping model proposed by Laughlin who attempted to explain the quantized Hall conductance in two dimensional systems\cite{22}. In this sense, one can immediately deduce that the cosine potential possesses no unique position, and can be replaced by arbitrary periodic potentials. They have the same topological structure.

\section*{5. Summary}
We have numerically computed the energy band structures of the discrete model and the continuous model as well. The Chern number is calculated with respective to the Bloch vector and the potential shift parameter. The adiabatic pumping of electrons driven by the shift of periodic potential is shown to be quantized and is equal to the topological invariants of the filled bands.

We thank W.A. Guo and H. Shao for helpful discussions. This work is supported by the NSF of China under grant No. 11374036 and the National 973 program under grant No. 2012CB821403.


\begin{thebibliography}{99}
\bibitem{1wen} X.-G. Wen, Adv. Phys. 44 (1995).
\bibitem{2Klitzing} K. Klitzing, G. Dorda, and M. Pepper, Phys. Rev. Lett. 45, 494 (1980).
\bibitem{3qi and zhang} X.-L. Qi and S.-C. Zhang, Rev. Mod. Phys. 83, 1057 (2011).
\bibitem{4Thouless} D. J. Thouless, M. Kohmoto, M. P. Nightingale, and M. den Nijs, Phys. Rev. Lett. 49, 405 (1982).

\bibitem{5Kane} M. Z. Hasan and C. L. Kane, Rev. Mod. Phys. 82, 3045 (2010).
\bibitem{6Kane} C. L. Kane and E. J. Mele, Phys. Rev. Lett. 95, 226801 (2005).
\bibitem{7Ber} B. A. Bernevig and S.-C. Zhang, Phys. Rev. Lett. 96, 106802 (2006).
\bibitem{8} B. A. Bernevig, T. L. Hughes, and S. C. Zhang, Science 314, 1757 (2006).
\bibitem{9} L. Fu and C. L. Kane, Phys. Rev. B 76, 045302 (2007).
\bibitem{10} T. Zhang, et al., Phys. Rev. Lett. 103, 266803 (2009).
\bibitem{11} L. Fallani, J. E. Lye, V. Guarrera, C. Fort, and M. Inguscio, Phys. Rev. Lett. 98, 130404 (2007).
\bibitem{12} T. Li, H. Kelkar, D. Medellin, and M. Raizen, Opt. Exp. 16, 5465 (2008).
\bibitem{13} L.-J. Lang, X. Cai, and S. Chen, Phys. Rev. Lett. 108, 220401 (2012).
\bibitem{14} Y. E. Kraus, Y. Lahini, Z. Ringel, M. Verbin, and O. Zilberberg, Phys. Rev. Lett. 109, 106402 (2012).
\bibitem{15} D. Hofstadter, Phys. Rev. B 14, 2239 (1976).
\bibitem{16} M. V. Berry, Proc. Royal Soc. London. A. Mathematical and Physical Sciences 392, 45 (1984).
\bibitem{17} T. Fukui, Y. Hatsugai, and H. Suzuki, arXiv preprint cond-mat/0503172  (2005).
\bibitem{18} Y. Hatsugai, arXiv preprint cond-mat/0405551 (2004).
\bibitem{19} Y. Hatsugai, Phys. Rev. B 48, 11851 (1993).
\bibitem{20} D. J. Thouless, Phys. Rev. B 27, 6083 (1983).
\bibitem{21} D. Xiao, M.-C. Chang, and Q. Niu, Rev. Mod. Phys. 82, 1959 (2010).
\bibitem{22} R. B. Laughlin, Phys. Rev. B 23, 5632 (1981).
\end{thebibliography}
\end{document}